\documentclass[aps,prl,nofootinbib,showpacs,floatfix,superscriptaddress,preprintnumbers]{revtex4}

 \usepackage{amsmath}
 \usepackage{graphicx}
\def\SM{$\mathrm{SU(3)_c \otimes SU(2)_L \otimes U(1)_Y}$ }
\def\vev#1{\left\langle #1\right\rangle}
\def\lfv{lepton flavour violation }

\newcommand{\AddrAHEP}{%
  AHEP Group, Institut de F\'{i}sica Corpuscular --
  C.S.I.C./Universitat de Val\`{e}ncia \\
  Edificio Institutos de Paterna, Apt 22085, E--46071 Valencia, Spain}

\newcommand{\AddrCFTP}{%
Departamento de F\'{i}sica and CFTP, Instituto Superior T\'{e}cnico \\
Technical University of Lisbon,1049-001 Lisboa, Portugal}

\newcommand{\AddrWurzburg}{%
  Institut f{\"u}r Theoretische Physik und Astrophysik,\\
  Universit{\"a}t W{\"u}rzburg, 97074 W{\"u}rzburg, Germany}

\relax

\begin{document}


\newcommand\T{\rule{0pt}{2.6ex}}       
\newcommand\B{\rule[-1.2ex]{0pt}{0pt}} 

\renewcommand{\Huge}{\Large}
\renewcommand{\LARGE}{\Large}
\renewcommand{\Large}{\large}
\def \znbb {$0\nu\beta\beta$ }
\def \nbb {$\beta\beta_{0\nu}$ }
\title{Neutrino mixing with revamped $A_4$ flavour symmetry}
\author{S.~Morisi}\email{ stefano.morisi@gmail.com} \affiliation{\AddrWurzburg}
\author{D.\,V.\,Forero}\email{dvanegas@ific.uv.es} \affiliation{\AddrCFTP }
\author{J.\,C.\,Rom\~{a}o}\email{jorge.romao@ist.utl.pt} \affiliation{\AddrCFTP} 
\author{J.~W.~F.~Valle} \email{valle@ific.uv.es} \affiliation{$^{1}$\AddrAHEP}


\begin{abstract}
  We suggest a minimal extension of the simplest $A_4$ flavour model
  that can induce a nonzero $\theta_{13}$ value, as required by recent
  neutrino oscillation data from reactors and accelerators.
  The predicted correlation between the atmospheric mixing angle
  $\theta_{23}$ and the magnitude of $\theta_{13}$ leads to an allowed
  region substantially smaller than indicated by neutrino oscillation
  global fits. Moreover, the scheme correlates CP violation in
  neutrino oscillations with the octant of the atmospheric mixing
  parameter $\theta_{23}$ in such a way that, for example, maximal
  mixing necessarily violates CP.  We briefly comment on other
  phenomenological features of the model.
 
\end{abstract}


\maketitle

\section{Introduction}
\label{intro}

The historic discovery of neutrino oscillations~\cite{art:2012}
constitutes a breakthrough in particle physics as it implies the need
of new physics beyond standard \SM model, the detailed nature of this
physics remains elusive, in particular regarding the flavour structure
of the the mechanism responsible for neutrino mass generation, and its
characteristic scale~\cite{Morisi:2012fg}.
Early studies establishing the oscillation phenomenon have indicated a
very specific pattern for the neutrino mixing angles, vastly different
from the CKM mixing
pattern~\cite{Maltoni:2004ei,schwetz:2008er}: while the atmospheric
angle $\theta_{23}$ is close to maximal, the solar angle $\theta_{12}$
is close to 30 degrees and no evidence was then present for a nonzero
$\theta_{13}$ value.
Although the latter may be accidental, it most likely follows a
rationale. This has motivated a strong effort towards the formulation
of symmetry–based approaches to address the flavour problem, in terms
of an underlying flavour symmetry of leptons and/or quarks, separately
or jointly. Indeed, these earlier observations were successfully
accounted for in terms of an underlying $A_4$ flavour
symmetry~\cite{Babu:2002dz,Altarelli:2005yp}.

However, recent accelerator experiments MINOS~\cite{Adamson:2011qu}
and T2K~\cite{Abe:2011sj,Abe:2013xua} as well as the measurements
reported by the Double CHOOZ~\cite{Abe:2011fz}, Daya
Bay~\cite{An:2012eh} and RENO reactor experiments~\cite{Ahn:2012nd}
have provided robust indications that $\theta_{13}$ is nonzero,
opening the door to the possibility of CP violation in neutrino
oscillations~\cite{nunokawa:2007qh,bandyopadhyay:2007kx}.
This finding provides a challenge for many $A_4$-based
schemes~\cite{Babu:2002dz,Altarelli:2005yp}, specially those leading
to the so--called tri-bimaximal (TBM) mixing ansatz proposed by
Harrison, Perkins and Scott~\cite{Harrison:2002er}. This scheme has
effective bimaximal mixing at the atmospheric scale and effective
trimaximal mixing at the solar scale.

Here we focus on the model was proposed by Babu, Ma and
Valle~\cite{Babu:2002dz} and studied in detail in \cite{hirsch:a4}. We
present a simple extension of the model that introduces an extra
scalar singlet flavon field $\zeta$ transforming as a $1^\prime$ of
$A_4$ to the Yukawa sector of the model.
We show explicitly how this breaks the remnant symmetry present in the
charged lepton sector~\footnote{ In Ref.\,\cite{Ma:2011yi} an extra
  scalar singlet was added in order to modify the mixing in the
  neutrino sector instead of the charged lepton sector. In contrast,
  deviations from the TBM ansatz may also arise from the charged
  lepton sector, as described
  in~\cite{Antusch:2011qg,Acosta:2012qf}.}, so as to induce a nonzero
$\theta_{13}$ value, hence making the model fully realistic and
opening the possibility of CP violation in neutrino oscillation.  Both
$\theta_{13}$ and the CP violation invariant $J_{CP}$ correlate with
the new term added to the model superpotential. In particular we show
how the model predicts a stringent correlation between the atmospheric
and the reactor mixing parameters, substantially more restrictive than
the allowed regions that emerge from recent global fits of neutrino
oscillations carried out within a generic flavour-blind scheme.  We
show how the model correlates CP violation in neutrino oscillation
with the octant of the atmospheric mixing parameter $\theta_{23}$, and
briefly comment on other possible phenomenological implications.

\section{The BMV model}
\label{sec:bmv-model}

We first recall the basic features of the Babu-Ma-Valle (BMV)
model~\cite{Babu:2002dz}.  The particle content is collected in
Tables~\ref{tab:t1} and~\ref{tab:t2}.
\begin{table}[!htb]
  \centering
\begin{tabular}{|c|c| c| c| c| c| c| } \hline
  & $\hat{Q}$ & $\hat{L}$ & $\hat{u}^c_1,\hat{d}^c_1,\hat{e}^c_1$ & $\hat{u}^c_2,\hat{d}^c_2,\hat{e}^c_2$ & $\hat{u}^c_3,\hat{d}^c_3,\hat{e}^c_3$ & $\hat{\phi}_{1,2}$ \T \B \\ \hline
   $A_4$& 3 &3 & $1$ & $1'$ & $1''$ & 1 \T \B \\ \hline
   $Z_3$& 1 & 1 & $\omega^2$ & $\omega^2$ & $\omega^2$ & 1 \T \B \\  \hline 
\end{tabular}
\caption{Flavour assignments of the MSSM fields, with $\omega=\exp{i2\pi/3}$.}
\label{tab:t1}
\end{table}
\begin{table}[!htb]
  \centering
\begin{tabular}{|c|c| c| c| c| c| c| c| c|} \hline
  & $\hat{U}$ & $\hat{U}^c$ & $\hat{D}$ & $\hat{D}^c$ & $\hat{E}$ & $\hat{E}^c$ & $\hat{N}^c$ & $\hat{\chi}$ \T \B \\ \hline
   $A_4$& 3 &3 & 3 &3 &3 &3 &3 &3 \T \B \\ \hline
   $Z_3$& 1 & 1 & 1 & 1 & 1 & 1 & 1 & $\omega$ \T \B \\  \hline 
\end{tabular} 
\caption{Mirror quark, lepton and Higgs superfield  assignments, with $\omega=\exp{i2\pi/3}$.}
\label{tab:t2}
\end{table}
The model implements an $A_4$ flavour symmetry within a supersymmetric
context.  $A_4$ is a discrete non-Abelian group of even permutations
of 4 objects, it has $\underline{1},\underline{1}',\underline{1}''$
and $\underline{3}$ irreducible representations (irrep) and it is the
smallest finite group with a triplet irrep.
The decomposition property of the product is:
\begin{equation}\label{eq:01}
\underline{3}\times \underline{3}=\underline{1}+\underline{1}'+\underline{1}''+\underline{3}+\underline{3}\,.
\end{equation}
The usual quark $\hat{Q}_i=(\hat{u}_i,\hat{d}_i)$, lepton
$\hat{L}_i=(\hat{\nu}_i,\hat{e}_i)$, and Higgs $\hat{\phi}_i$
transforms under $A_4$ as given in Table~\ref{tab:t1}.  In addition
one adds the heavy quark, lepton, and Higgs superfields indicated in
Table~\ref{tab:t2}. These are all $SU(2)$ singlets.

The superpotential of the BMV model is then given by:\\
\small{
\begin{equation}\label{eq:02}
\begin{split}
\hat{W}&= M_U\hat{U}_i\hat{U}^c_i+f_u \hat{Q}_i\hat{U}^c_i\hat{\phi}_2+h_{ijk}^u\hat{U}_i\hat{u}^c_i\hat{\chi}_k\\
       &+ M_D\hat{D}_i\hat{D}^c_i+f_d \hat{Q}_i\hat{D}^c_i\hat{\phi}_2+h_{ijk}^d\hat{D}_i\hat{d}^c_i\hat{\chi}_k\\
       &+M_E\hat{E}_i\hat{E}^c_i+f_e\hat{L}_i\hat{E}^c_i\hat{\phi}_1+h_{ijk}^e\hat{E}_i\hat{e}^c_j\hat{\chi}_k\\
       &+ \frac{1}{2}M_N\hat{N}_i^c\hat{N}^c_i+f_N\hat{L}_i\hat{N}^c_i\hat{\phi}_2 +\mu \hat{\phi}_1\hat{\phi}_2\\
       &+\frac{1}{2}M_{\chi}\hat{\chi}_i\hat{\chi}_i+h_\chi \hat{\chi}_1\hat{\chi}_2\hat{\chi}_3\,.
\end{split}
\end{equation}
}
The scalar potential involving $\chi_i$ is given by:
\begin{equation}\label{eq:03}
V=|M_\chi \chi_1 +h_\chi \chi_2 \chi_3|^2+|M_\chi \chi_2 +h_\chi \chi_3 \chi_1|^2+|M_\chi \chi_3 +h_\chi \chi_1 \chi_2|^2,
\end{equation}
which have the supersymmetric solution ($V=0$)
\begin{equation}\label{eq:04}
\langle \chi_1 \rangle=\langle \chi_2 \rangle=\langle \chi_3 \rangle=u
\end{equation}
We assume that the $A_4$ flavour symmetry is broken softly at some high
scale~\cite{Babu:2002dz}.

The Dirac mass matrix linking $(e_i,E_i)$ to $(e_j^c,E_j^c)$ can be
written as:
\begin{equation}\label{eq:05}
  \mathcal{M}_{e E} =
  \begin{bmatrix}
    0 & 0 & 0 & f_ev_1 & 0 & 0\\
    0 & 0 & 0 & 0 & f_ev_1 & 0\\
    0 & 0 & 0 & 0 & 0 & f_ev_1\\
    h_1^e\,u & h_2^e\,u & h_3^e\,u & M_E & 0 & 0\\
    h_1^e\,u & h_2^e\,u\,\omega & h_3^e\,u \,\omega^2& 0 & M_E & 0\\
    h_1^e\,u & h_2^e\,u\,\omega^2 & h_3^e\,u\,\omega & 0 & 0 & M_E
  \end{bmatrix}
  \equiv
  \begin{bmatrix}
    0 & X_1^D \\
    X_2 & Y^D\\
  \end{bmatrix}\,,
\end{equation}
where $v_1 = \langle \phi_1^0\rangle$\,\footnote{Here $\phi_{1,2}$ are the usual two Higgs of supersymmetry.}, with similar forms also for the
corresponding quark mass matrices.  After block diagonalization of
Eq. (\ref{eq:05}), one finds that the reduced $3\times 3$ Dirac mass
matrix for the charged leptons is diagonalized by the magic matrix
$U_\omega$:
\begin{equation}\label{eq:06}
U_\omega
   = \frac{1}{\sqrt{3}}
  \begin{bmatrix}
    1&1&1\\
    1&\omega&\omega^2\\
    1&\omega^2&\omega\\
  \end{bmatrix}\,.
\end{equation}
For $f_e v_1 \ll h_i\,u \ll M_E$ the charged lepton masses are
obtained as
\begin{equation}\label{eq:07}
   \tilde{m}_i^2\simeq \frac{3 f_e^2 v_1^2 }{M_E^2} \frac{h_i^{e\,2} u^2}{1 + 3 (h_i^e u)^2/M_E^2}\,.
\end{equation}

Turning to the neutral sector, the Majorana mass matrix in the basis
$(\nu_i,N_i^c)$ and in the basis where charged leptons are diagonal, is given by:
\begin{equation}\label{eq:08}
\mathcal{M}_{\nu N}=
 \begin{bmatrix}
 0 & f_N v_2\,U_\omega\\
 f_N v_2\,U_\omega^T & M_N
 \end{bmatrix}\,,
\end{equation}
where $v_2=\langle\phi_2^0\rangle$. Hence, the reduced light neutrino
mass matrix after seesaw becomes:
\begin{equation}\label{eq:09}
\mathcal{M}_{\nu}=\frac{f_N^2\,v_2^2}{M_N} U_\omega^T U_\omega=m_0
 \begin{bmatrix}
 1 & 0 & 0\\
 0 & 0 & 1\\
 0 & 1 & 0
 \end{bmatrix}
=m_0\,\mathbf{\lambda}\,.
\end{equation}
leading to degenerate neutrino masses at this stage. Eq. (\ref{eq:09})
is corrected by the wave function renormalizations of $\nu_i$, as well
as the corresponding vertex renormalizations \cite{Babu:2002dz}. Given
the structure of the $\lambda_{ij}$ at the high scale
(Eq. (\ref{eq:09})), its form at low scale is fixed to first order as:
\begin{equation}\label{eq:10}
\mathbf{\lambda}=
\begin{bmatrix}
    1+2\delta_{ee} & \delta_{e\mu}+\delta_{e\tau} & \delta_{e\mu}+\delta_{e\tau}\\
    \delta_{e\mu}+\delta_{e\tau} & 2 \delta_{\mu \tau} & 1+\delta_{\mu \mu}+\delta_{\tau \tau}\\
    \delta_{e\mu}+\delta_{e\tau} & 1+\delta_{\mu \mu}+\delta_{\tau \tau} & 2 \delta_{\mu \tau}
  \end{bmatrix}\,,
\end{equation}
where all parameters are assumed to be real \cite{Babu:2002dz}.
Rewriting  Eq.\,(\ref{eq:09}) with $\delta_{0} \equiv \delta_{\mu
  \mu}+\delta_{\tau \tau}-2\delta_{\mu \tau}$, $\delta \equiv 2\delta_{\mu \tau}$, 
$\delta^\prime \equiv \delta_{ee}-\delta_{\mu\mu}/2-\delta_{\tau \tau}/2$ and 
$\delta^{\prime \prime} \equiv \delta_{e\mu}+\delta_{e\tau} $ one has
\begin{equation}\label{eq:11}
\begin{bmatrix}
    1+\delta_{0}+2\delta+2\delta^\prime & \delta^{\prime \prime} & \delta^{\prime \prime}\\
    \delta^{\prime \prime} & \delta & 1+\delta_{0}+\delta\\
    \delta^{\prime \prime} & 1+\delta_{0}+\delta & \delta
  \end{bmatrix},
\end{equation}
so that the eigenvectors and eigenvalues can be determined
\textit{exactly}. The effective neutrino mixing matrix is given by
\begin{equation}\label{eq:12}
U_{\nu}(\theta) = 
  \begin{bmatrix}
    \cos \theta & -\sin \theta & 0\\
     \sin \theta/\sqrt{2}& \cos \theta/\sqrt{2} & -1/\sqrt{2}\\
    \sin \theta/\sqrt{2} & \cos \theta/\sqrt{2} & 1/\sqrt{2}\\
  \end{bmatrix},
\end{equation}
while the three light neutrino mass eigenvalues are
\begin{equation}\label{eq:13}
\begin{split}
\lambda_1&=1+\delta_0+2\delta+\delta^\prime-\sqrt{\delta^{\prime \,2}+2\delta^{\prime \prime\,2}}\\
\lambda_2&=1+\delta_0+2\delta+\delta^\prime+\sqrt{\delta^{\prime \,2}+2\delta^{\prime \prime\,2}}\\
\lambda_3&=-1-\delta_0
\end{split}
\end{equation}
so that one finds the BMV model predictions for the neutrino
for the mixing angles, given as
\begin{equation}\label{eq:14}
\begin{split}
& \tan^2 \theta_{12}=\frac{\delta^{\prime \prime\,2} }{\delta^{\prime \prime\,2}+\delta^{\prime \,2}-\delta^\prime \sqrt{\delta^{\prime \,2}+2\delta^{\prime \prime\,2} } }\\
& \sin^2 \theta_{13}=0 \\
& \tan^2 \theta_{23}=1 \Rightarrow \text{maximal}
\end{split}
\end{equation}
For the other oscillation parameters, namely the squared mass square
differences, assuming $\delta^\prime,\delta^{\prime \prime}
\ll\delta$, one has
\begin{equation}\label{eq:15}
\begin{split}
& \Delta m^2_{31}\simeq \Delta m^2_{32} \simeq 4\delta \,\,m_0^2 \\
& \Delta m^2_{21}\simeq 4\sqrt{\delta^{\prime \,2}+2\delta^{\prime \prime\,2} }\, m_0^2
\end{split}
\end{equation}
One sees that the mixing matrix in the neutrino sector in
Eq. (\ref{eq:14}) has just one free parameter $\theta$ which
corresponds to the unpredicted solar mixing angle $\theta_{12}$.
One now assumes that radiative corrections lift the neutrino mass
degeneracy, as required by the solar neutrino oscillation data.
Using the solar angle in Eq. (\ref{eq:14}) and the square mass
differences Eq. (\ref{eq:15}), one can estimate the size of some of
the wave function and vertex corrections required in order to fit the
observed oscillation parameters. One finds the following relations
\begin{equation}\label{eq:16}
\begin{split}
\frac{\delta}{|\delta^\prime|}&=\xi \frac{\Delta m_{31}^{2}}{\Delta
  m_{21}^{2}} \left(\frac{1}{1-2\, \sin^2{\theta}} \right)\approx 92.96\, \xi,\\
\frac{|\delta^{\prime\,\prime}|}{|\delta^\prime|}
&=\sqrt{\frac{1}{2}\left[\left(\frac{1}{1-2\, \sin^2{\theta}}
  \right)^2 -1\right]}\approx 1.83,
\end{split}
\end{equation}
where $\xi=1(-1)$ correspond to the case of
$\delta^\prime<0\,(\delta^\prime>0)$. In order to fit neutrino
oscillation data, the threshold parameter $\delta^\prime$ must be of
the same order as $\delta^{\prime\,\prime}$ and also
$\delta^{\prime},\delta^{\prime\,\prime} \ll \delta$. With
$\delta^{\prime }<0$ and $|\delta^{\prime \prime}/\delta^{\prime
}|=1.8$ the predicted neutrino mixing pattern is indeed consistent
with the oscillation data before the latest T2K, Daya Bay and RENO
results for $\theta_{13}$.

\section{Revamping the original $A_4$ model}
\label{sec:revamping-model}

The main goal of this paper is to accommodate the current neutrino
data~\cite{Tortola:2012te} within a minimally extended $A_4$-based BMV
scenario.
In general, the effective mixing in the leptonic sector is given by:
\begin{equation}\label{eq:17}
K=U_\nu(\theta),
\end{equation}
where we have rotated by the magic matrix $U_\omega$. The idea is now
to generate modifications of the mixing in the leptonic sector
$U_{\omega}'=U_{\omega}\,U_\delta$, in such a way that the
modified lepton mixing matrix is now given by
\begin{equation}\label{eq:18}
K'=U_\delta^\dagger\, U_\nu(\theta)
\end{equation}
where $U_\delta$ denotes a correction which may yield a
nonvanishing $\theta_{13}$ while keeping good predictions for the
other neutrino oscillation parameters, in particular, the atmospheric
mixing angle $\theta_{23}$.

\subsection{Charged lepton corrections to lepton mixing}
\label{sec:charg-lept-corr}

As a first attempt we relax the condition used to obtain the charged
lepton masses, Eq.~(\ref{eq:07}), by allowing the $M_E$ scale (see
Eq.~(\ref{eq:05})) to lie at the TeV scale~\footnote{This would lead
  to the existence of flavour-changing neutral currents at the tree
  level. These would induce sizeable lepton flavour violating
  processes. }. This results in unitarity violation corrections to the
lepton mixing matrix.
With $M_E$ in Eq. (\ref{eq:05}) at the TeV scale, one must take into
account not only the first order terms in the block diagonalization of
the mass matrix of the charged lepton sector as in Eq. (\ref{eq:07})
but also the next to leading order effects.
Using the Schechter-Valle procedure~\cite{Schechter:1982cv} for the
block diagonalization one finds
\begin{equation}\label{eq:19}
U =\mathcal{U}\cdot V=\exp(iH)\cdot V\, \qquad
H= 
\left(
\begin{array}{cc}
0& S\\
S^\dagger & 0
\end{array}
\right),
\qquad V=
\left(
\begin{array}{cc}
V_1&0\\
0 & V_2
\end{array}
\right),
\end{equation}
where $H$ is an anti-Hermitian operator and $V_i$ are unitary matrices
which diagonalize each block. The $S$ matrix is determined at first
order by the diagonalization condition
$U^\dagger\,M\,U=\text{Diag}\{m_i\}$ for a given Hermitian matrix $M$.
In our specific case:
\begin{equation}\label{eq:20}
 \mathcal{M}_{eE} (\mathcal{M}_{eE})^\dagger= 
\begin{bmatrix}
    (f_ev_1)^2\,I& M_E\,f_ev_1\,I\\
     M_E\,f_ev_1\,I& U_\omega ( \text{Diag}\{3(h_i^eu)^2\})U_\omega^\dagger+M_E^2\,I
  \end{bmatrix}
  \equiv
  \begin{bmatrix}
    m_1^2 & m_2^2\\
    m_2^{2 \,\dagger} & m_3^2
  \end{bmatrix}
 \end{equation}
the $S$ is given by:
\begin{equation}\label{eq:21}
i\,S=-m_2^2(m_1^2-m_3^2)^{-1}=U_\omega \,\text{diag}\{-M_E\,f_ev_1 [(f_ev_1)^2-3(h_i^eu)^2-M_E^2]^{-1}\}\,U_\omega^\dagger
\end{equation}
where the first term in Eq (\ref{eq:20}) and second in Eq (\ref{eq:21})
correspond to our specific case.

In order to calculate the next to the leading order terms, one expands
the exponential in Eq.~(\ref{eq:19}) in a power series in $S$. The
next to the leading order terms are combinations of the Identity and
products of $S~S^\dagger$ and $S$. Given the structure of the $S$
matrix in Eq.~(\ref{eq:21}) is clear that even if we go to higher
orders in the expansion, the effective charged lepton mass will always
be diagonalized by the magic matrix $U_\omega$. In other words
$U_\delta\equiv 1$.

The origin of the structure of the $S$ matrix in Eq. (\ref{eq:21})
comes from the fact that in the BMV model, the matrices in the upper
right corner and the lower right corner in Eq. (\ref{eq:05}) are
proportional to the identity.
The net effect is that, even allowing for unitarity violation in the
charged sector, does not change the structure of the lepton mixing
matrix. Somehow a remnant symmetry of the $A_4$ remains that leads to
$\theta_{13}\equiv 0$.

\subsection{Minimal flavon extension of the original $A_4$ model}
\label{sec:minim-flav-extens}

In order to break the unwanted remnant symmetry present in the charged
lepton sector of the model, we now add a scalar singlet flavon field
$\zeta$ to the superpotential in Eq. (\ref{eq:02}). The flavon scalar
field $\zeta$ transforms as a $1^\prime$ under the $A_4$ flavour
symmetry. This leads to a new superpotential term of the form:
\begin{equation}\label{eq:22}
\zeta (E\,E^c)_{1''}
\end{equation}
where we will parametrize the flavon scale as $\vev{ \zeta}=\beta\, M_E$.
This results in a new mass matrix for the lower right corner of
Eq. (\ref{eq:05}) that now has the structure:
\begin{equation}\label{eq:23}
Y_D=M_E \times I + \beta M_E \times \text{Diag}\{1,\omega,\omega^2\},
\end{equation}
so that the corresponding charged lepton matrix in Eq. (\ref{eq:20})
is now given by
\begin{equation}\label{eq:24}
 \mathcal{M}_{eE} (\mathcal{M}_{eE})^\dagger = 
  \begin{bmatrix}
    (f_ev_1)^2\,I& f_ev_1\,Y_D^\dagger\\
     f_ev_1\,Y_D & U_\omega
     (\text{Diag}\{3(h_i^eu)^2\})U_\omega^\dagger + Y_DY_D^\dagger
  \end{bmatrix}
\end{equation}
where $Y_D$ is no longer diagonalized by the magic matrix. This
changes structure of the $S$ matrix in Eq. (\ref{eq:21}) and breaks
the unwanted remnant symmetry which lead to $\theta_{13}\equiv 0$.
As a consequence, one obtains a corrected matrix $U_{\omega}^c$ that
leads to a $U_\delta$ matrix of the form
\begin{equation}\label{eq:25}
U_\omega'=U_\omega \, U_\delta,
\end{equation}
with which the effective lepton mixing matrix from Eq. (\ref{eq:17})
can be calculated.  The modified lepton mixing $K'$ is a complex
non-unitary 3x3 matrix from which one must extract the three angles
and three CP phases that characterize the simplest neutrino mixing
parameter set.  One finds that, indeed, the proposed flavon extension
of the original $A_4$ model scheme can engender a nonzero value for
the reactor mixing angle, as required by recent neutrino oscillation
data.

\section{Neutrino oscillation parameters}
\label{sec:neutr-oscill-param}

\begin{figure}[!htb]
\centerline{
\includegraphics[scale=0.5]{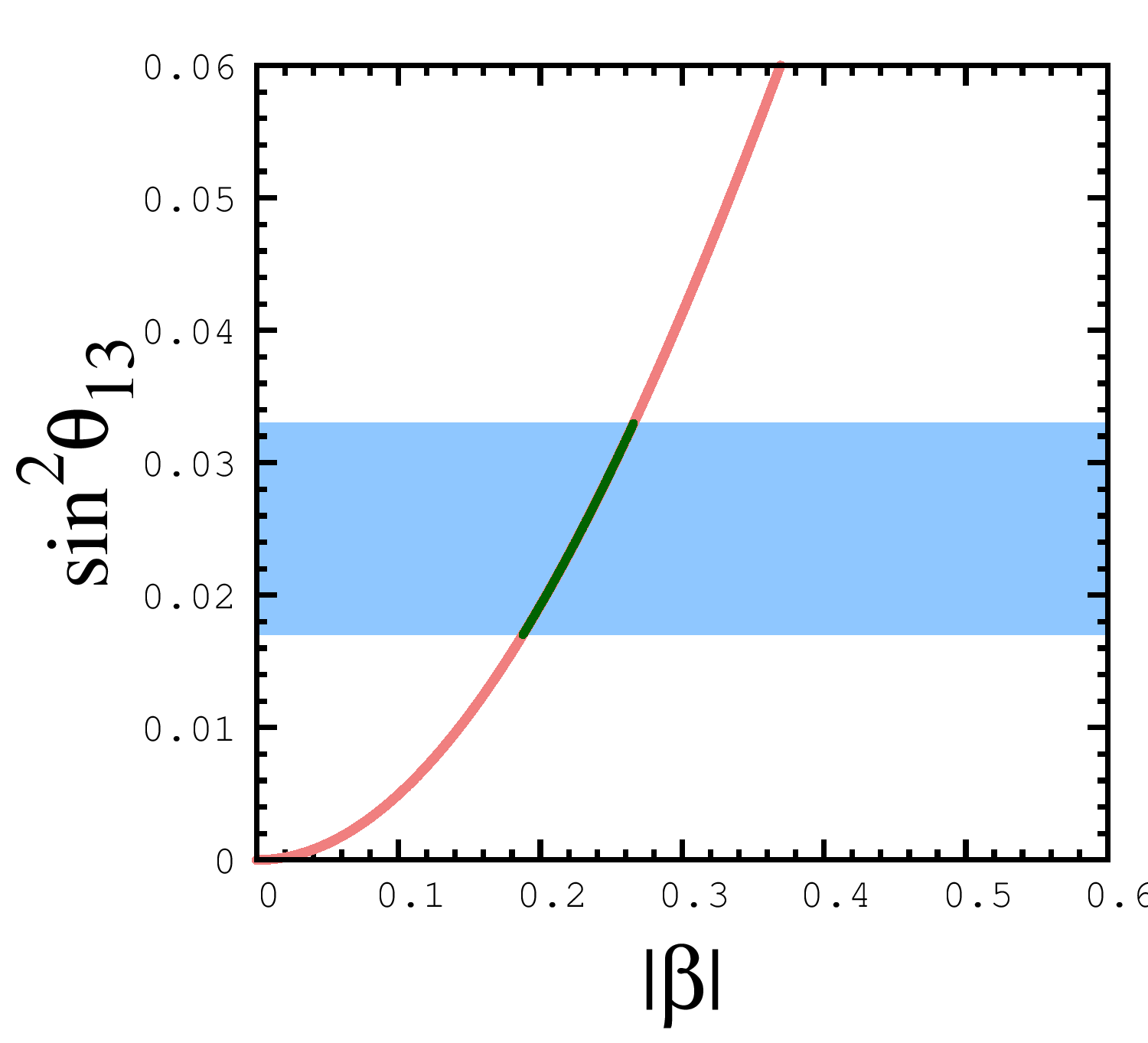}
\includegraphics[scale=0.5]{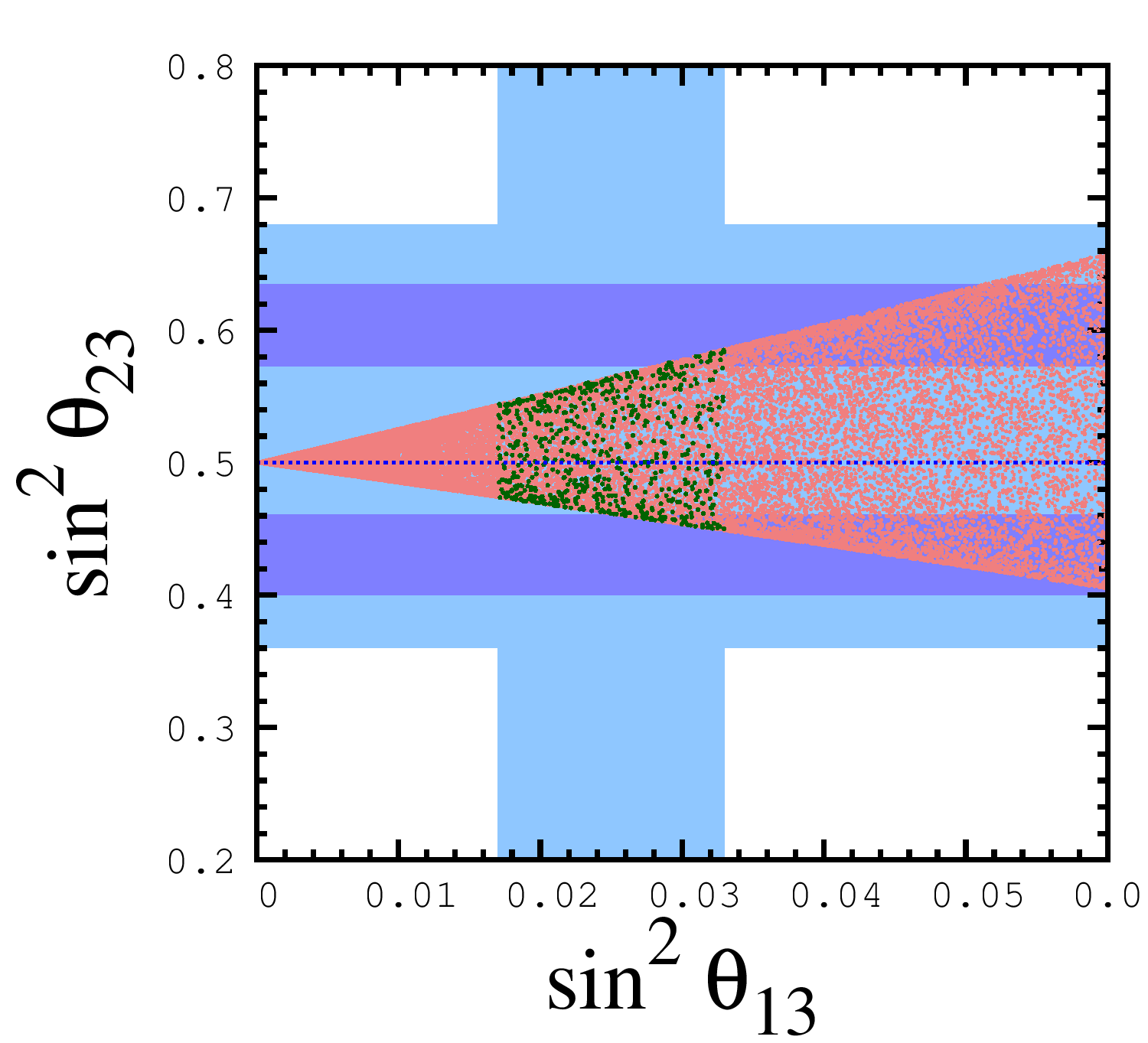}
}
\caption{\label{fig:2} (Left) Correlation between the reactor angle
  $\sin^2{\theta_{13}}$ and the magnitude of the flavon coupling
  parameter $|\beta|$. (Right) The ``triangle'' region gives the
  predicted correlation between atmospheric and reactor angles for
  different $\beta$ parameter choices. The broad vertical (horizontal)
  bands are the current allowed values for $\sin^2{\theta_{13}}$
  ($\sin^2{\theta_{23}}$) at $3\,\sigma$. In both panels the flavon
  phase $\phi_\beta$ has been varied continuously in the range
  $-\pi/2\le \phi_\beta \le \pi/2$. All points in the ``triangle'' are
  allowed by the $\theta_{12}$ $3\,\sigma$ solar angle range, but only
  the green (dark) points are consistent with $\theta_{13}$ as
  well. Finally the two thin horizontal bands correspond to the
  $1\,\sigma$ preferred regions in the global oscillation fit
  of~\cite{Tortola:2012te}.}
\end{figure}
\begin{figure}[!b]
\centerline{
\includegraphics[scale=0.5]{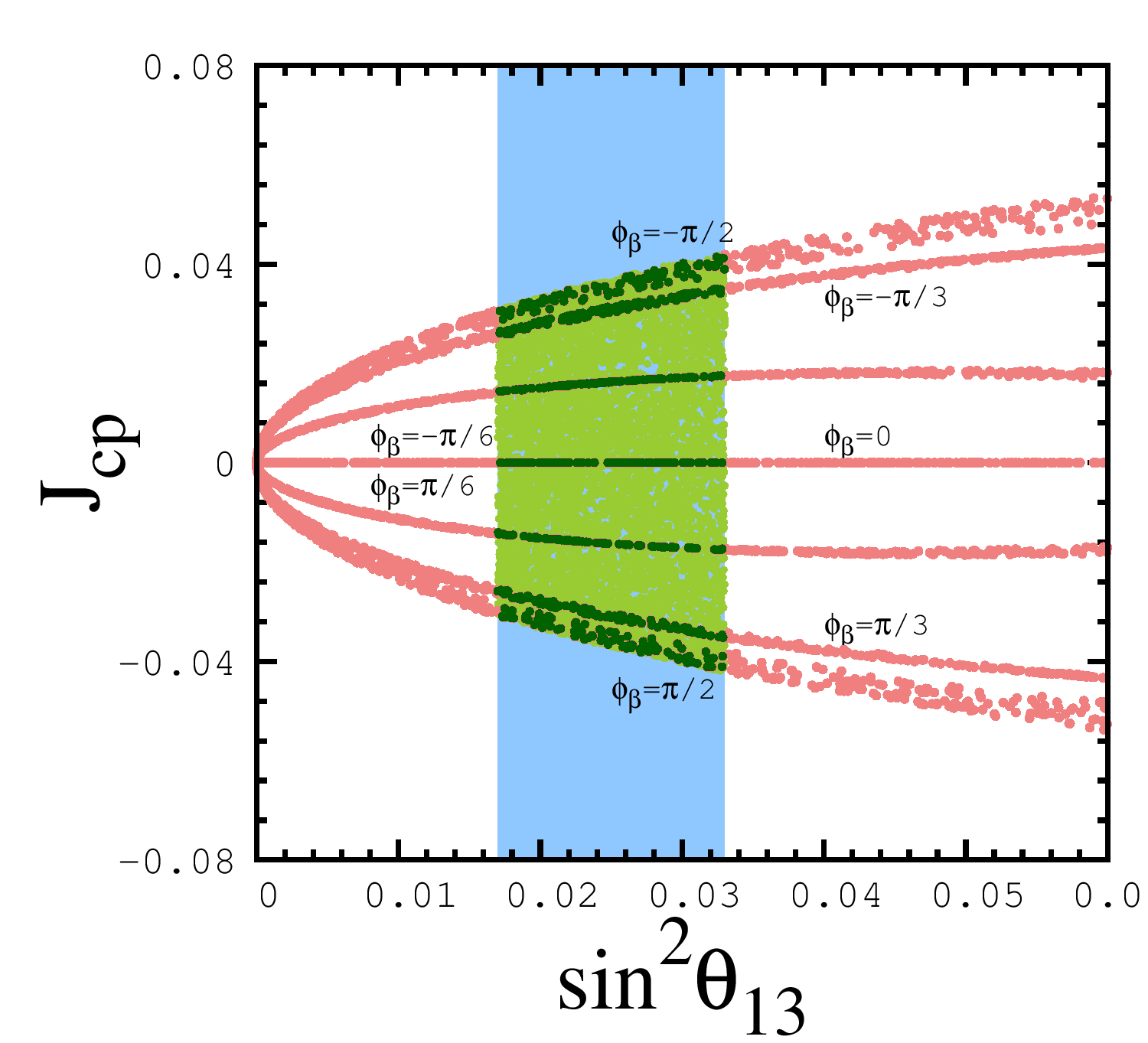}
\includegraphics[scale=0.5]{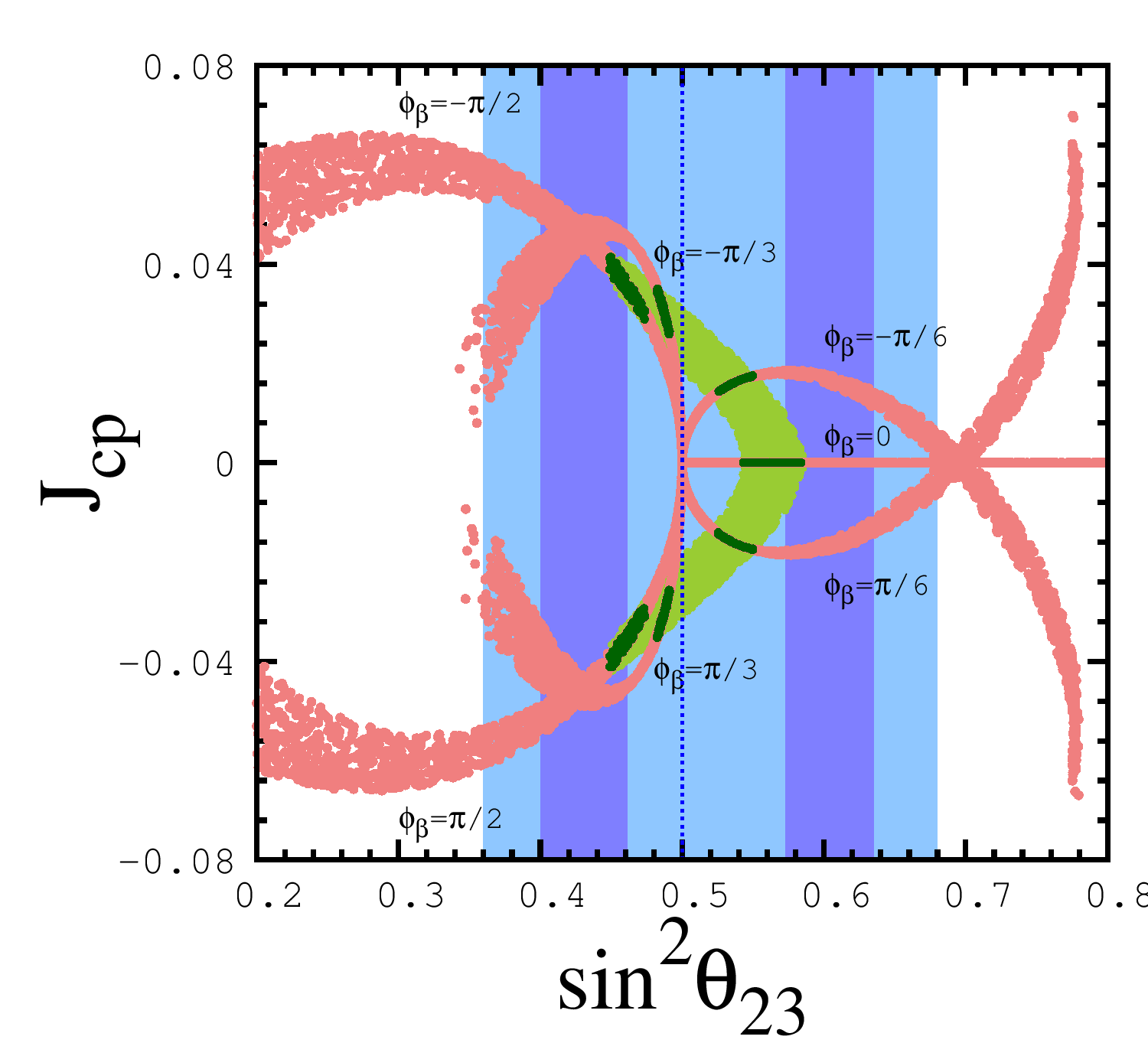}
}
\caption{\label{fig:3} Correlation between the magnitude of the CP
  violation invariant $J_{CP}$ and the two mixing angles, reactor
  $\theta_{13}$ and atmospheric $\theta_{23}$, in left and right
  panels, respectively. Discrete set of phase values have been used in
  the range $-\pi/2 \le \phi_\beta \le \pi/2$ in steps of
  $\pi/6$. Random points in pink are compatible with the current
  $3\,\sigma$ range of the solar angle $\theta_{12}$, but only the
  green points are compatible with the $\theta_{13}$ and $\theta_{23}$
  range at $3\,\sigma$.  }
\end{figure}

Using the modifications to the BMV model explained above
(e.g. Eq. (\ref{eq:24})), we have obtained quantitative results by
numerically diagonalized the charged lepton mixing matrix in
Eq.~(\ref{eq:05}).  The three mixing angles, are obtained directly as
\begin{equation}\label{eq:26}
  \begin{split}
  \tan \theta_{12}&=|K'_{1,2}(\theta)|/|K'_{1,1}(\theta)|,\\
 \sin \theta_{13}&=|K'_{1,3}(\theta)|,\\
 \tan \theta_{23}&=|K'_{2,3}(\theta)|/|K'_{3,3}(\theta)|,\\
   \end{split}
\end{equation}
where the $\theta$ parameter has been varied randomly in the range
$0\le \sin^2 \theta \le 1$. The scales $f_e v_1$ and $M_E$ have also
been varied randomly in the range $1 \le f_e v_1 \le 10^2$ GeV and
$10^4 \le M_E \le 10^5$ GeV, leading to the results presented in
Fig. \ref{fig:2}. As we will see for such values the mixing matrix
$K'$ is well described by a unitary approximation.

As one can see from the left plot of Fig. \ref{fig:2}, in order to
generate a non vanishing reactor angle $\theta_{13}$ the magnitude in
the flavon coupling $|\beta|$ must be nonzero. In principle this
result is independent of the phase $\phi_\beta$.

On the other hand from the right panel of Fig. \ref{fig:2}, one sees
how the new coupling engenders not only a nonzero $\theta_{13}$ value,
but also a restricted range for the atmospheric angle $\theta_{23}$.
If one takes at face value the hints for non-maximal $\theta_{23}$ at
$1\,\sigma$ which follow from global oscillation
fits~\cite{Tortola:2012te} then one finds that the allowed regions for
$\theta_{23}$ in each octant would be very narrow indeed. However
currently maximal atmospheric mixing remains perfectly
consistent~\cite{Abe:2013xua}.  As one sees in Fig.~\ref{fig:3} for
maximal atmospheric mixing, the flavon phase must have a non zero
value, as seen in the right panel of Fig.~\ref{fig:3}.

Note that the allowed region is modulated by the value of the $\beta$
phase $\phi_\beta$, in other words, as one varies the values of the
phase $-\pi/2 \le \phi_\beta \le \pi/2$ one sweeps the
triangle--shaped region indicated in the right panel of
Fig.~\ref{fig:2}. One finds a linear correlation between the ``opening
angle'' of the triangle and the magnitude of the continuous phase
angle $\phi_\beta$. Intermediate $\phi_\beta$ values cover the
indicated shaded sub-region of the vertical strip.
While all current neutrino mixing angles, including the reactor angle
$\theta_{13}$, are consistent with a real flavon coupling, allowing
the latter to be complex results in a determination of the octant for
$\theta_{23}$ as shown in the right panel of Fig. \ref{fig:3}. A
measurement of the violating CP phase would imply a determination of
the octant, or vice versa. Again, continuous phase values in between
the extremes lead to the half-moon-like region indicated in the right
panel in Fig.~\ref{fig:3}.



In other to further clarify the issue of leptonic CP violation within
this model, we now turn to the Dirac phase $\delta_{CP}$ associated to
CP violation in neutrino oscillations.
Rather than trying to extract this phase directly, we have calculated
the associated Jarlskog parameter $J_{CP}$
\begin{equation}\label{eq:27}
J_{CP}=\mathcal{I}\{K_{e1}^*K_{\mu 3}^*K_{e3}\,K_{\mu 1}\},
\end{equation}
which is invariant under any conceivable phase redefinitions. Our
numerical result is shown in Fig. \ref{fig:3}, in which we have
numerically evaluated Eq. (\ref{eq:27}) for a discrete values of the
phase $-\pi/2 \le \phi_\beta \le \pi/2$ in steps of $\pi/6$. One can
see that for $|\beta|>0$ the invariant $J_{CP}$ is non zero in
correlation with the non zero value of the phase $\phi_\beta$.
%
%
By allowing the flavon coupling $\beta$ to be complex one not only
introduces CP violation in neutrino oscillations, but also selects the
allowed octant of the atmospheric mixing angle $\theta_{23}$ in
correspondence with the assumed values of the phase $\phi_\beta$,
which is clearly seen from right panel of Fig. \ref{fig:3}. This
constitutes an important prediction of the model which may be tested
in the future neutrino oscillation experiments.
In contrast the Majorana phases can hardly be probed within this model
since the mass spectrum is almost degenerate, so that there can never
be an important destructive interference between different \znbb
amplitudes. As a result the \znbb decay rate is expected to be large
and should be probed in current and future experiments.

\section{Analytical understanding}
\label{sec:analyt-underts}

In order to gain a better understanding of the proposed scheme, we now
turn to an analytic approach. We have fixed the $M_E$ scale to be
$10^2$ times bigger than the TeV scale and we have obtained the
correlations already displayed in Fig. \ref{fig:2}. The result in the
left panel suggests a simple theoretical relation. Indeed, assuming
$K'$ to be nearly unitary, we are within a perturbative limit where
we can solve the problem analytically, by diagonalizing the effective
charged lepton mass matrix at the leading order and keeping only the
terms until second order in $|\beta|$. This way we find a simple
approximate result for the reactor angle given as
\begin{equation}\label{eq:28}
\sin^2 \theta_{13}=|\beta|^2 \frac{h_1^8-2 h_1^6\,h_3^2+2 h_1^4\,h_3^4-2 h_1^2\,h_2^2\,h_3^4+h_2^4\,h_3^4-2 h_1^2\,h_3^2(h_1^2-h_2^2 )(h_1^2-h_3^2 )\cos{2\,\phi_\beta}}{2 [(h_1^2-h_2^2 )(h_1^2-h_3^2 )]^2}
\end{equation}
in terms of the Yukawa parameters $h_i$ that determine the charged
lepton masses through Eq. (\ref{eq:07}).
\begin{figure}[!b]
\centerline{
\includegraphics[scale=0.45]{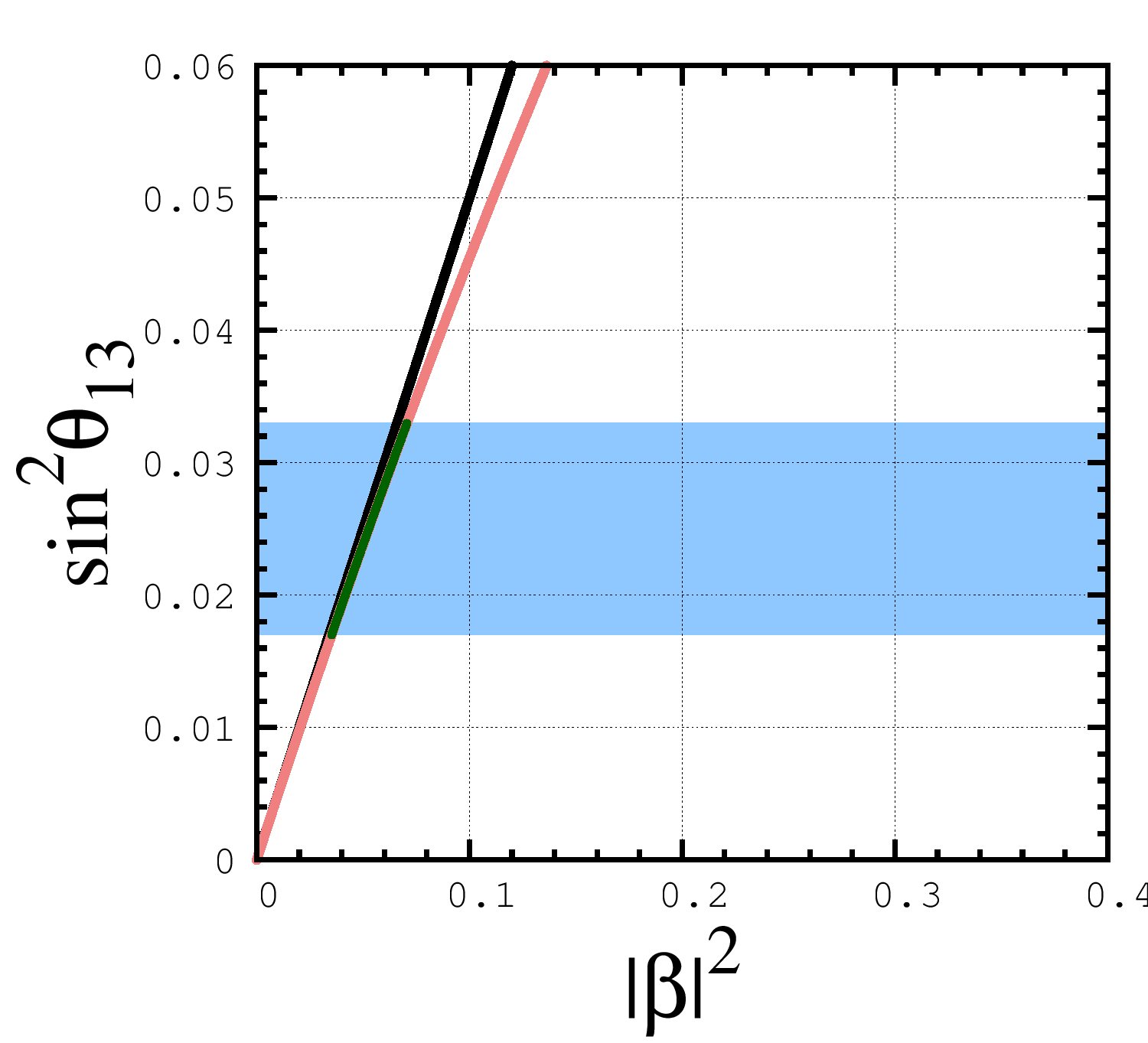}
\includegraphics[scale=0.45]{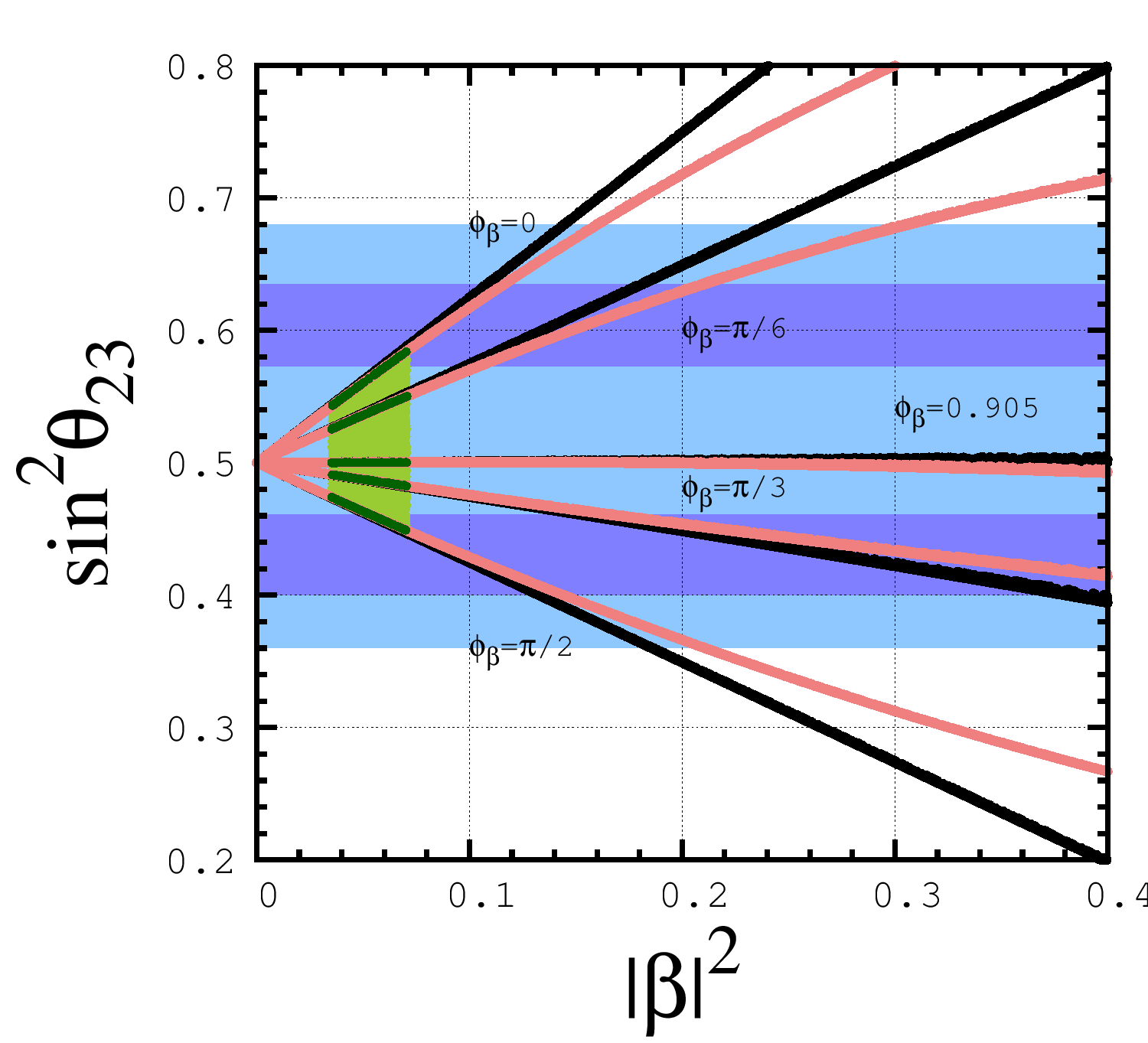}
\includegraphics[scale=0.45]{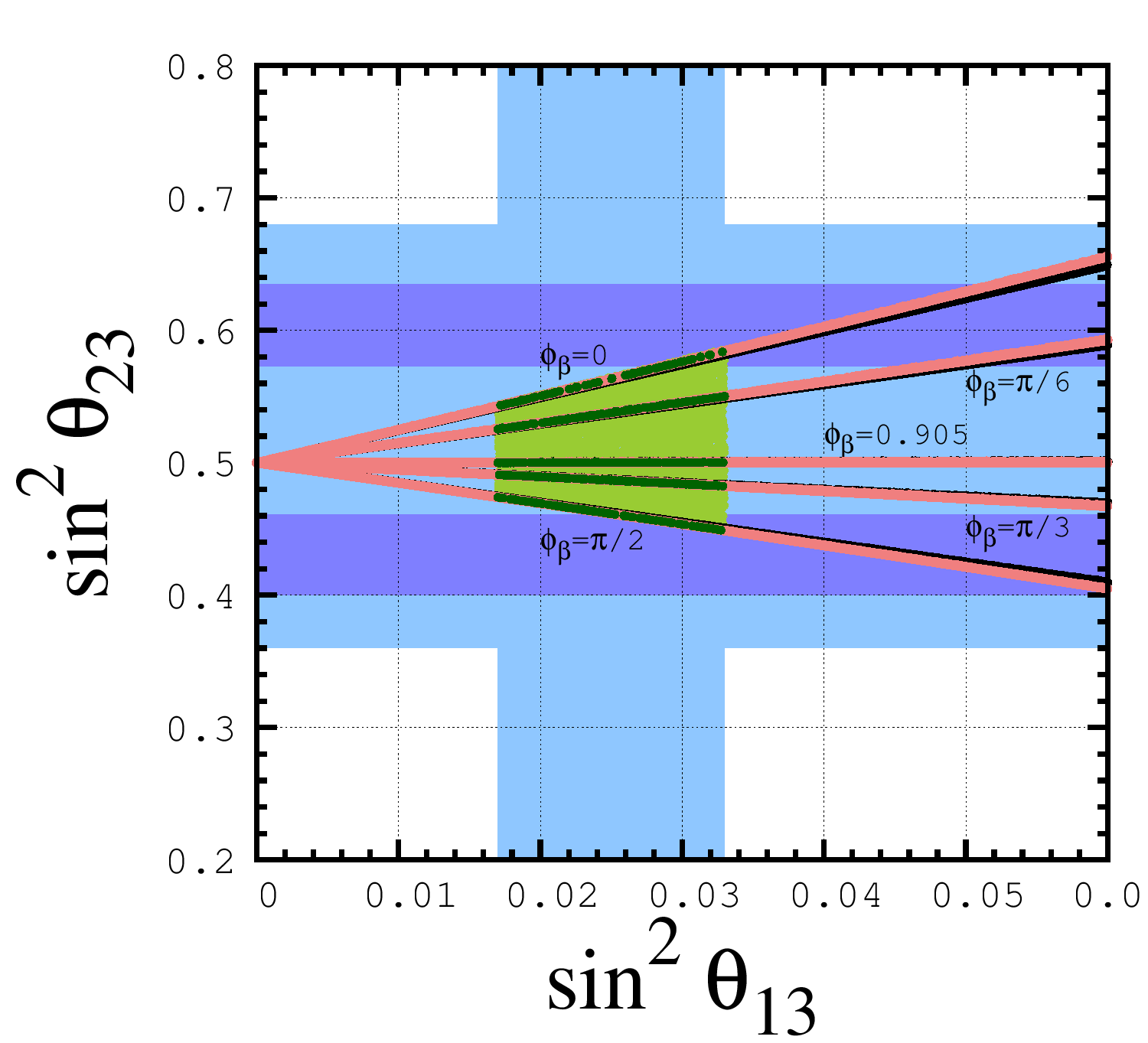}
}
\caption{\label{fig:4} Here we give the exact numerically determined
  predictions for the reactor and atmospheric mixing parameters
  $\theta_{13}$ $\theta_{23}$ in terms of the magnitude of $\beta$ and
  its phase $\phi_\beta$ varied in steps of $\pi/6$. We also give the
  results that follow from the approximate expressions in
  Eqs. (\ref{eq:28}),(\ref{eq:29}). Numerical results are in pink
  while analytical ones are in black.  There is rather good agreement
  within the currently allowed $3\sigma$ range of the neutrino mixing
  angles: $\theta_{12}$, $\theta_{13}$ and $\theta_{23}$ indicated by
  the blue bands. Notice that the negative values of the flavon phase
  corresponds to the same correlation, which is more clear from
  Eqs. (\ref{eq:28}),(\ref{eq:29}).  }
\end{figure}
This way one can explain analytically the right panel of
Fig. \ref{fig:2} and conclude that $\sin \theta_{13}$ can be non
vanishing even if the value of the $\beta$ phase is zero.
In a completely analogous procedure, we have also obtained an
approximate relation for the atmospheric angle
\begin{equation}\label{eq:29}
\begin{split}
\sin^2 \theta_{23}&=\frac{1}{2}+|\beta| \frac{h_2^2}{h_2^2-h_3^2} \cos{\phi_\beta} \\
&+|\beta|^2 \frac{(-h_1^8+2 h_1^6\,h_3^2+h_2^4\,h_3^4)(h_2^2-h_3^2)+2 h_1^2(h_2^2\,h_3^6-h_2^4\,h_3^4)+2 h_3^2(h_1^2-h_2^2 )(h_1^2-h_3^2 )[h_1^2(h_2^2+h_3^2)-2 h_2^2\,h_3^2] \cos{2\phi_\beta} }{4 [(h_1^2-h_2^2 )(h_1^2-h_3^2 )]^2 (h_2^2-h_3^2)}.
\end{split}
\end{equation}
Numerically we have checked that the expansions in $|\beta|$ leading
to the expressions in Eq. (\ref{eq:28}) and Eq. (\ref{eq:29})
reproduce very well the numerical results for the correlations such
as, for instance, those given by the curve in left panel of
Fig. \ref{fig:2} within the current allowed range indicated by global
neutrino oscillation fits and summarized by the blue bands displayed
in Fig. \ref{fig:4}.  As we have already noted, for special values of
$\phi_\beta$ the octant of $\theta_{23}$ gets determined as shown in
Fig. \ref{fig:2} and Fig. \ref{fig:4}.

Before concluding let us make one last comment on the size of the
corrections in the neutrino sector.  Within the revamped BMV model we
have now introduced, the mixing predictions have been recalculated
through the square mass differences are given by Eq. (\ref{eq:15}). As
we have already mentioned, the free parameter $\theta$ in the mixing
corresponds to the solar angle $\theta_{12}$ for a given values of the
underlying radiative corrections. Due to the modified neutrino mixing
pattern the correspondence between the free parameter and the solar
angle through the radiative corrections that come from the soft
symmetry breaking sector and, strictly speaking, these are no longer
the same as in the original flavon-less BMV model.

\section{Discussion}
\label{sec:discussion}

We have proposed a minimal extension of the simplest $A_4$ flavour
model of Babu, Ma and Valle that can induce a nonzero $\theta_{13}$
value, as required by recent neutrino oscillation data coming from
reactors and accelerators.
We have shown how the predicted correlation between the atmospheric
mixing angle $\theta_{23}$ and the magnitude of $\theta_{13}$ leads to
an allowed region that is substantially smaller than indicated by
model-independent neutrino oscillation global fits. Moreover, our
proposed scheme establishes a correlation between CP violation in
neutrino oscillations and the octant of the atmospheric mixing
parameter $\theta_{23}$.  In particular one finds that, for example,
maximal atmospheric mixing as well as the first octant necessarily
violate CP. Currently we find that both are consistent at the
1$\sigma$ level with the global (including atmospheric data) neutrino
oscillation analysis of Ref.~\cite{Tortola:2012te}.  We also stress
that ours is a quasi-degenerate neutrino scenario.  Recent
restrictions on the absolute neutrino mass from the Planck
collaboration~\cite{2013arXiv1303.5076P} indicate values for the
parameter $\delta$ characterizing slepton radiative corrections for
which \lfv induced by supersymmetric particle exchanges is expected to
lie at the limits. That would provide another complementary way to
probe this model. This issue will be taken up elsewhere.

\section{Acknowledgments}

Work supported by MINECO grants FPA2011-22975, and MULTIDARK
Consolider CSD2009-00064, by Prometeo/2009/091 (Gen.  Valenciana), by
EU ITN UNILHC PITN-GA-2009-237920. S.M. thanks DFG grant WI
2639/4-1. J.C.R. also acknowledges the financial support from grants
CFTP-FCT UNIT 777, CERN/FP/123580/2011 and PTDC/FIS/102120/2008.

\bibliographystyle{h-physrev4} 

\end{document}